\documentclass[usenatbib,conference]{basi}
\usepackage[T1]{fontenc}
\usepackage[british]{babel}
\usepackage[varg]{txfonts}
\usepackage{rotating}
\usepackage{dcolumn}
\begin{document}
\title[The Deep Full-Stokes Radio Sky]{The Deep Full-Stokes Radio Sky}
\author[A.~R.~Taylor et.~al.]%
       {A.~R.~Taylor$^{1,2}$\thanks{email: \texttt{artaylor@ucalgary.ca}},
       S.~Bhatnagar$^3$, J.~Condon$^3$, D.~A.~ Green$^4$, J.~M.~Stil$^{5}$,
       \newauthor P.~Jagannathan$^5$, N.~Kantharia$^6$, R.~Kothes$^7$, 
       R.~Perley$^3$, J.~Wall$^8$, T.~Willis$^7$\\
       $^1$Department of Astronomy, University of Cape Town, Cape Town, South Africa\\
       $^2$Department of Physics, University of the Western Cape, Cape Town, South Africa\\
       $^3$National Radio Astronomy Observatory, USA\\
       $^4$ Cavendish Laboratory, University of Cambridge, Cambridge, UK\\
       $^5$Department of Physics and Astronomy, University of Calgary, Calgary, Canada\\
       $^6$ National Centre for Radio Astrophysics, Pune, India\\
       $^7$ Dominion Radio Astrophysical Observatory, National Research Council, Penticton, Canada\\
       $^8$ Department of Physics and Astronomy, University of British Columbia, Vancouver, Canada}

\pubyear{2014}
\volume{00}
\pagerange{\pageref{firstpage}--\pageref{lastpage}}

\date{Received --- ; accepted ---}

\maketitle
\label{firstpage}

\begin{abstract}
The new broad-band capabilities of large radio interferometers such as the GMRT and JVLA allow for 
long-integration mosaic imaging observations to create ultra-deep full-polarization images of the sky 
over wide frequency ranges.  
Achieving rms sensitivities of order 1\,$\mu$Jy, these observations explore the radio source population at 
flux densities well below the regime dominated by classical radio galaxies and Active Galactic Nuclei. 
We present initial results from radio sources revealed with deep mosaicking
observations with the GMRT and JVLA at respectively 0.6 and 5 GHz, and evidence that the $\mu$Jy sensitivity 
level marks the transition to detection of polarized emission from a population of
sources dominated by emission from magnetic fields in the disks of starburst and normal
galaxies.

\end{abstract}


\section{Introduction}\label{s:intro}

Magnetic fields are an essential part of many astrophysical phenomena, but fundamental questions remain 
about their evolution, structure and origin. 
Exploring the magnetic universe is one of the key science themes of the Square Kilometre Array project
\citep{Schilizzi}.
Polarimetry of radio synchrotron emission is one of the most successful probes of magnetic fields.
Imaging the deep polarized sky will be required to probe the evolution and role of cosmic magnetism
over a significant fraction of cosmic history.  

\cite{Taylor07} and  \cite{Grant10} first studied populations with polarized flux densities below 
1 mJy at  1.4 GHz.  At these levels the polarized population remains dominated by
radio galaxies and AGN with total flux densities of 10's of mJy.
A key goal of deep polarization imaging is the detection of polarized emission from galaxy disks. 
The integrated fractional polarization of a sample of nearby disk galaxies was measured at 4.8 GHz 
by \cite{Stil09}, who showed that at least 60\% of normal spirals are polarized at higher than 1\%.
Radiation from more distant galaxies begins to become a significant fraction of the radio wavelength source population 
below a total intensity flux density of a few mJy (e.g.\  Condon 1984).
Detection of polarized emission from galaxy disks in the distant universe thus requires imaging observations
capable of detecting down to $\mu$Jy levels and below.
In this paper we provide an initial report on deep polarization imaging observations with the GMRT and JVLA
to begin the exploration of the $\mu$Jy polarized sky.


\section{Observations and Imaging}\label{s:obs}

GMRT observations were carried in several sessions from  2011 to 2013. 
An area of 1.2 sq degrees of ELAIS N1 was covered by a mosaic of 7 pointings arranged
in a hexagon pattern centred on $\alpha$ = 16$^{\rm h}$ 10$^{\rm m}$ 30$^{\rm s}$,  
$\delta$ = 54$^{\rm o}$ 35 00$''$.    Each pointing was observed for
approximately 30 hours in three 10-hour sessions. 
Data were taken in 256 spectral channels in four polarization states covering 
a 32 MHz bandwidth centred at 612 MHz. 
 The flux scale, bandpass and absolute position angle calibration was 
secured by observations of 3C286 twice in each observing session.  Time dependent
gains and on-axis polarization leakage corrections were measured by frequent 
observations of J1549+506.    To reduce the noise and mitigate the effect of off-axis gain 
and polarization response in the central regions of the mosaic, the separation of the 
pointings  on the sky was closely spaced at 16$'$, or 38\% of the FWHM.

JVLA observations of a smaller, 0.13 sq degree, region centred on the GMRT field were obtained in
the C and B configurations during 2012 and 2013, using 3C286 and J1624+5652 for primary and secondary
polarization calibration.   
Each of 10 closely-spaced (3.5$'$ or 35\% of the FWHM) pointings was observed for about 3 hours 
in each array configuration, for a total of 60 hours.  Data were taken in 1024 channels over a 2 GHz band
centred at 5 GHz. Additional observation in A configuration are ongoing.

\begin{figure}
\centerline{\includegraphics[width=8cm]{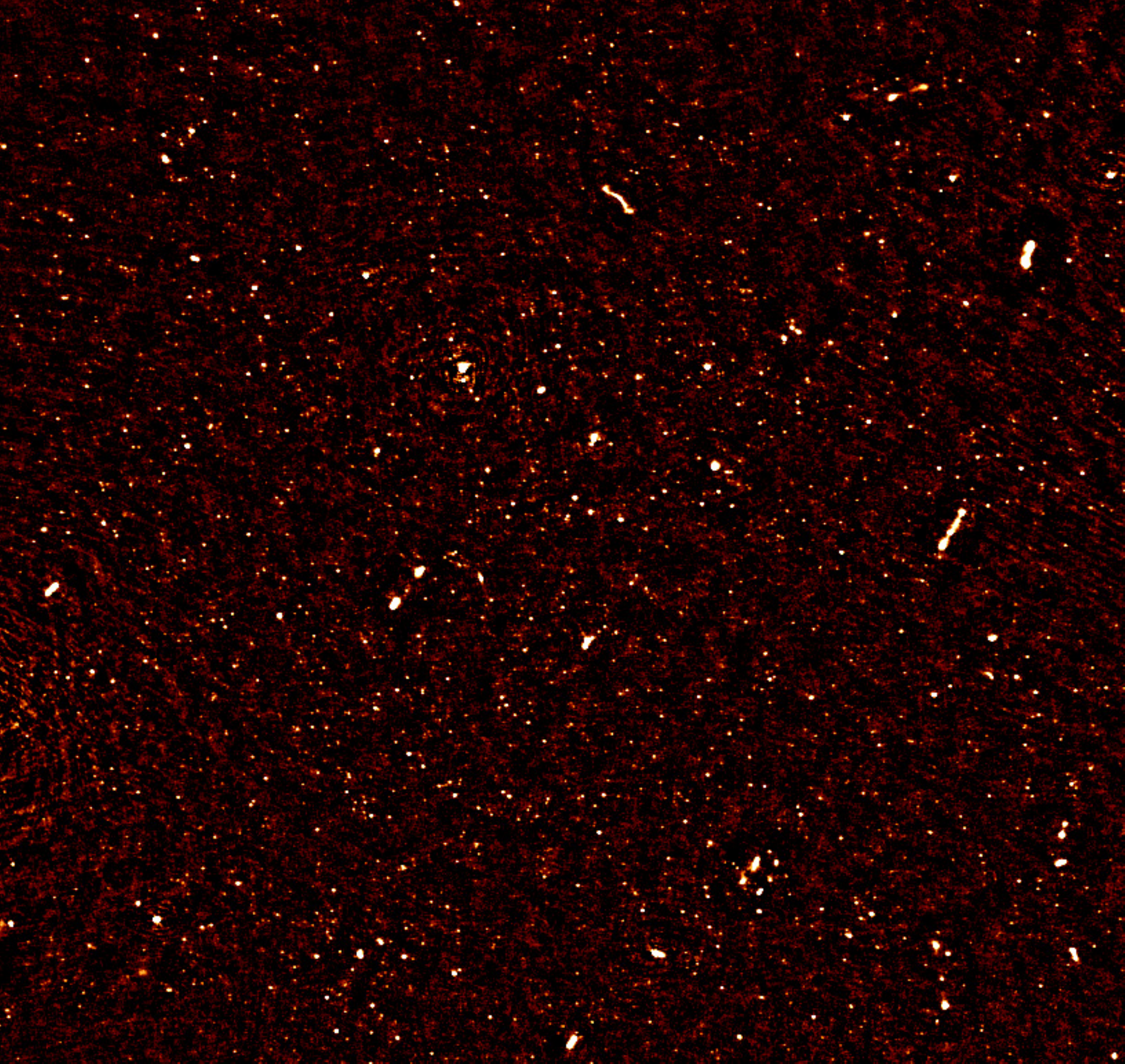}}
\caption{Stokes I image of the inner 0.9 sq deg of
the GMRT ELAIS N1 mosaic at 612 MHz.\label{fig:gmrt_total}}
\end{figure}

\begin{figure}
\centerline{\includegraphics[width=8cm]{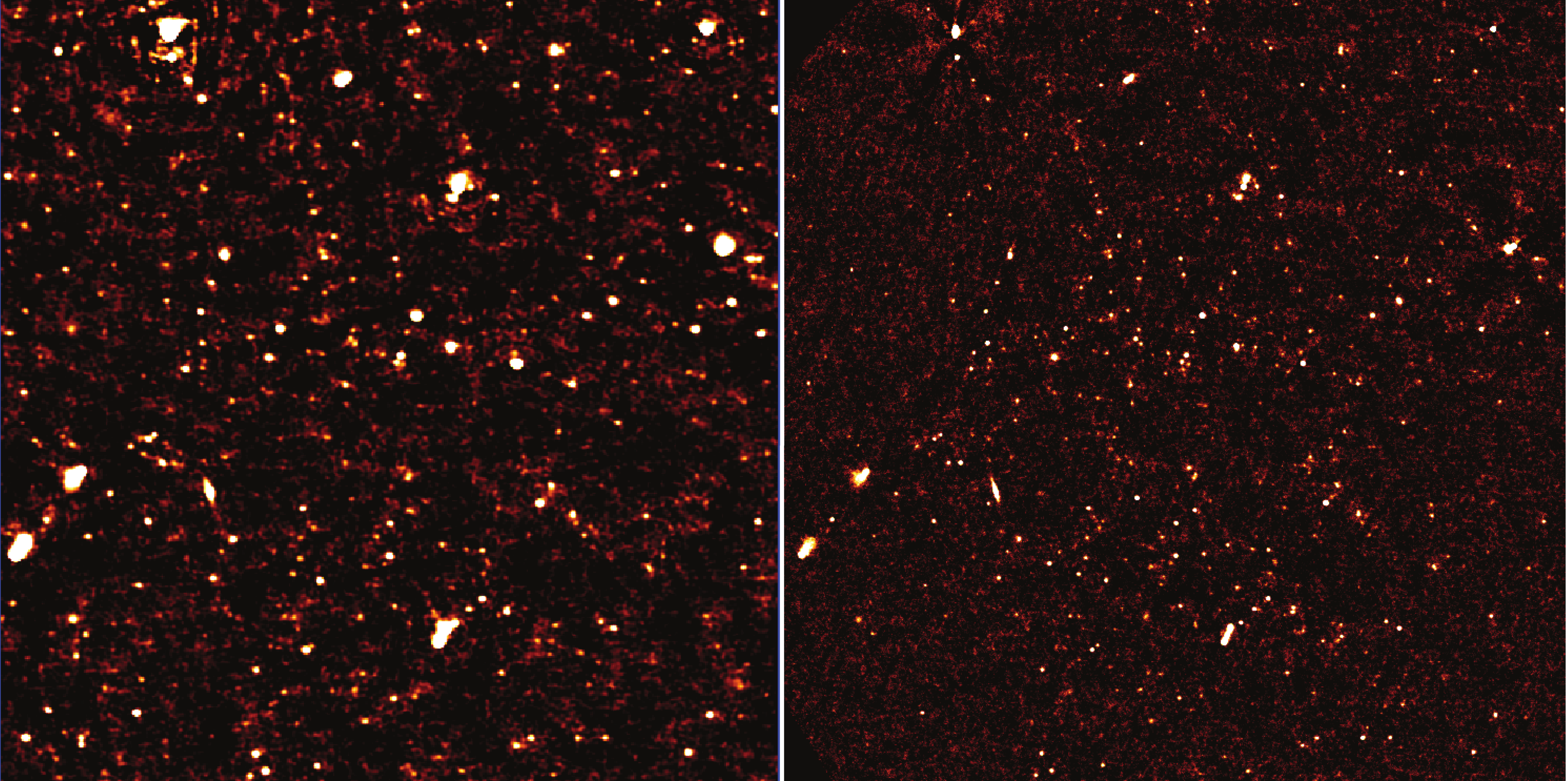}}
\caption{Stokes I image of the inner 0.1 square degrees of
the GMRT ELAIS N1 mosaic at left and the JVLA 5 GHz mosaic image of the same
area at right.  
\label{fig:gmrt_jvla_total}}
\end{figure}

Both GMRT and JVLA visibilities were calibrated, imaged and mosaicked using the CASA processing
software.  The central 0.9 sq degrees of the GMRT mosaic image are shown in Figure~\ref{fig:gmrt_total}.
The rms noise in this mosaic is 10.3\,$\mu$Jy/beam before primary beam correction.   The angular resolution is 6.1$'' \times 5.1''$.
Within the region defined by the half-power point of the 6 outer pointings, there
are 2800 sources above a flux density of 50\,$\mu$Jy.     The JVLA 5 GHz mosaic image is shown
at right in Figure~\ref{fig:gmrt_jvla_total}.  The angular resolution of the JVLA image is 2.5$''$ and the
rms noise in this image is 1.05\,$\mu$Jy/beam before primary beam correction.
Within an area of 0.12 square degrees 483 radio sources are detected above a flux density of 5\,$\mu$Jy.

\begin{figure}
\begin{center}
\begin{tabular}{p{7.2cm}cp{3.8cm}}
\raisebox{-\height} {\includegraphics[width=8.5cm]{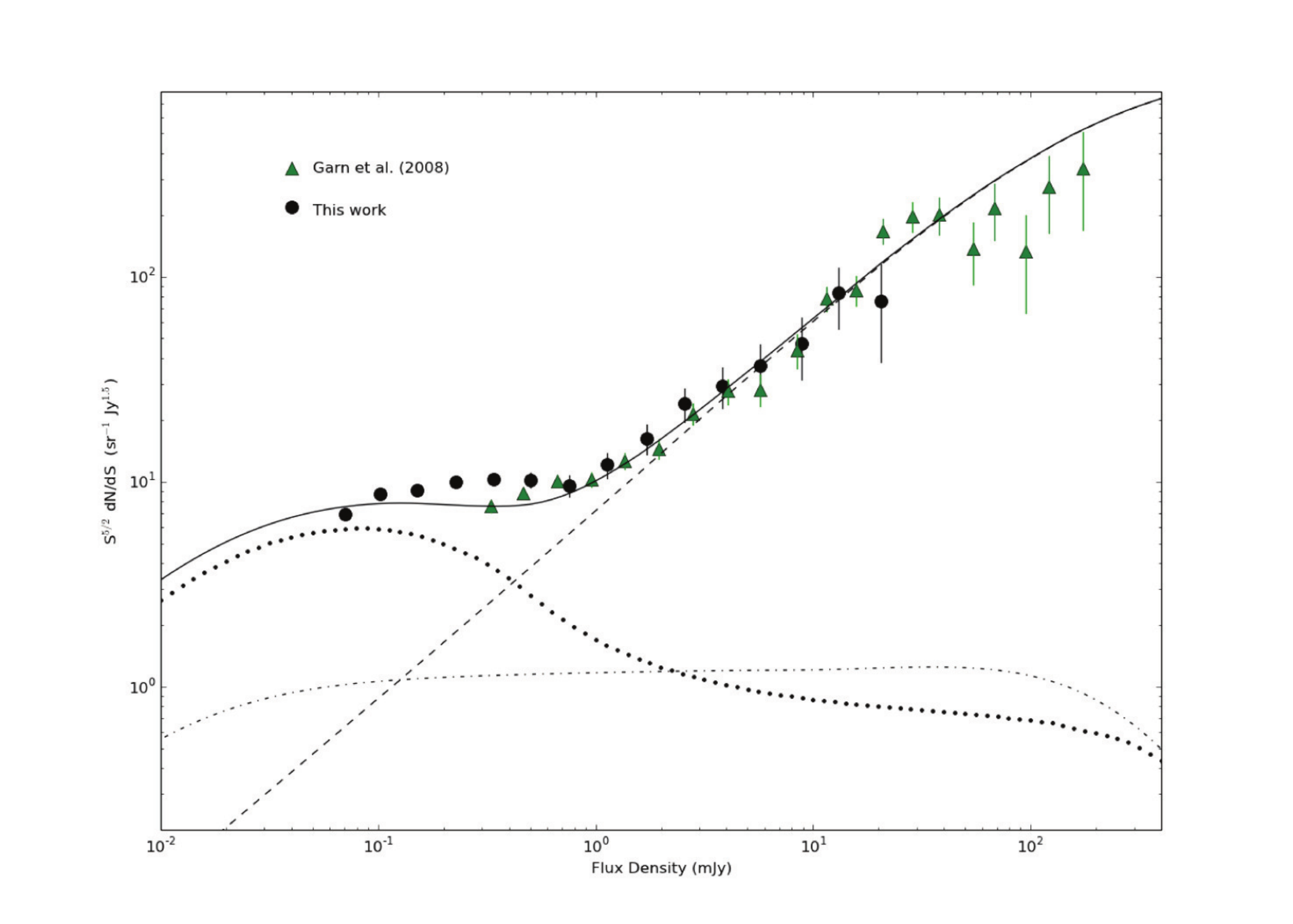}} & \quad &
\caption{The total intensity differential source count at 612 MHz.  Solid black circles are counts from
this work to a flux density limit of 70\,$\mu$Jy.  Triangles show counts 
from \cite{Garn}.  Curves are model source counts from \cite{Massardi}, 
for AGN (dashed), starburst galaxies (dotted) and normal galaxies (dot-dashed).
The solid line is the total counts from all three populations.
\label{fig:counts}}
\end{tabular}
\end{center}
\end{figure}

The GMRT image shows a handful of classical bright double-lobed radio galaxies.  However the majority  of
the sources form a dense ``background'' of compact sources scattered throughout the image that are 
largely unresolved at 6$''$ resolution. The Stokes $I$ 
differential source counts derived from the mosaic are shown in Figure~\ref{fig:counts}.
 The counts flatten  at about 1\,mJy and continue approximately flat down to the detection limit.  
 Simulated  counts by \cite{Massardi}, also shown on the figure, indicate 
that below about 450\,$\mu$Jy, the population is dominated by star forming galaxies.   
A smaller fraction of more nearby "normal" galaxies is expected to dominate over AGN at around
100\,$\mu$Jy.

\section{Polarization of the $\mu$Jy Sky}

\begin{figure}
\centerline{\includegraphics[width=8cm]{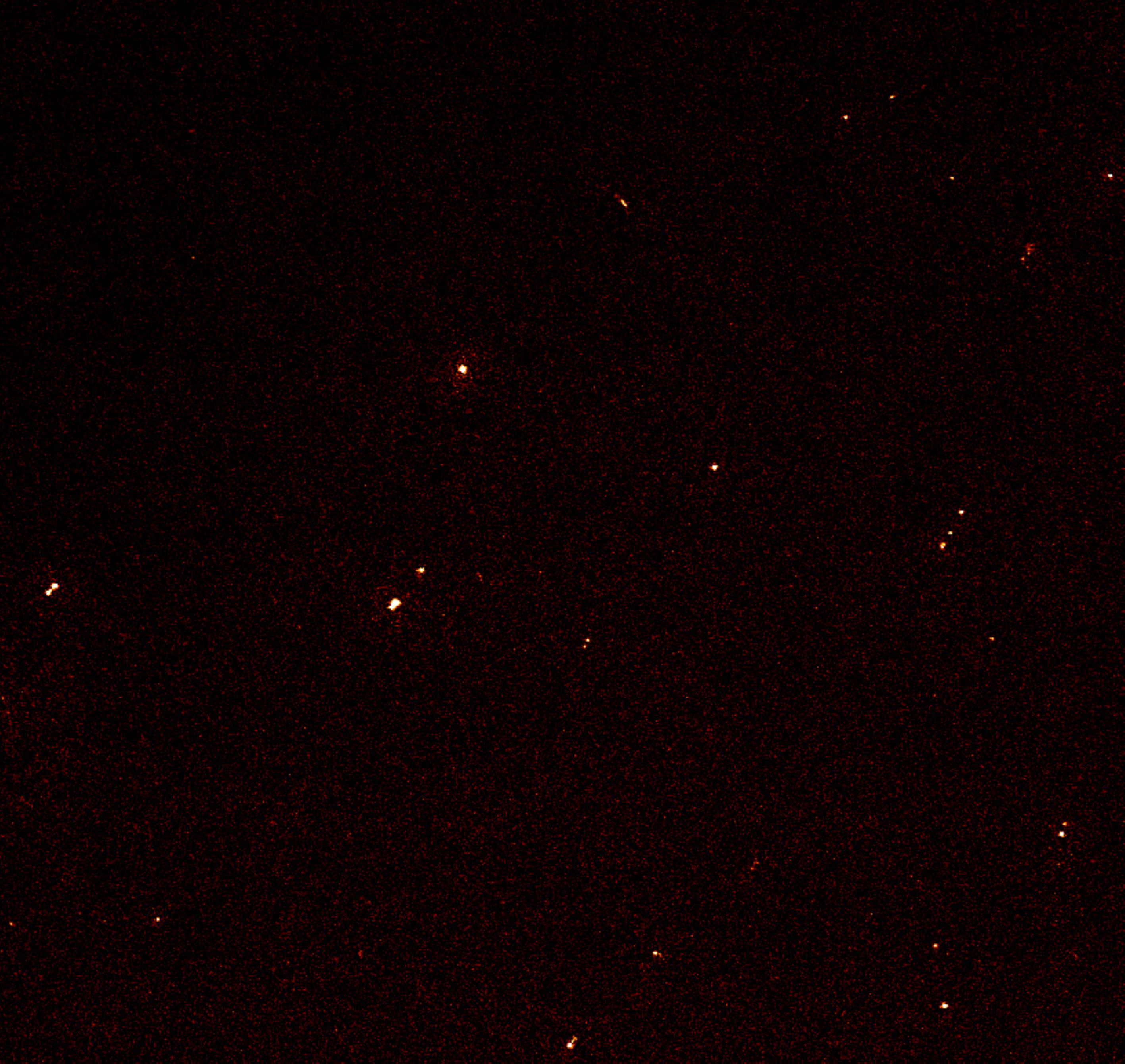}}
\caption{GMRT 612 MHz polarized intensity mosaic image of the region shown
in Figure~\ref{fig:gmrt_total}. 
\label{fig:gmrt_pol}}
\end{figure}

\begin{figure}
\centerline{\includegraphics[width=10cm]{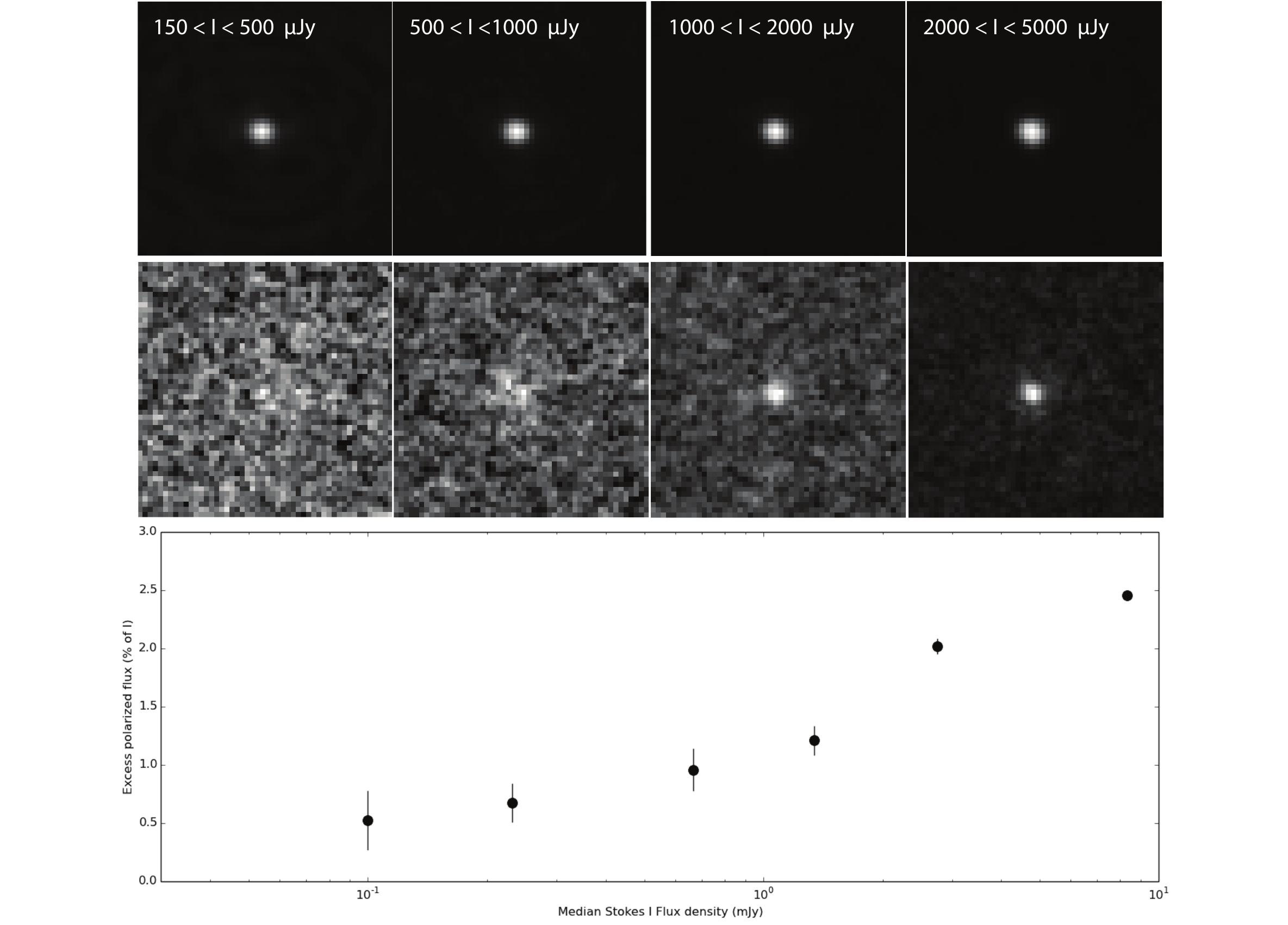}}
\caption{Stacking sources at 612 MHz.  Top row are median total
intensity images in four stacking bins range from 150\,$\mu$Jy to
5000 \,$\mu$Jy.   Below are stacked median polarized intensity images.
The bottom panel shows observed percent polarization excess above
the background in the stacked median images as a function of Stokes I
flux density.  \label{fig:polstack}}
\end{figure}

The rms noise in the GMRT Stokes $Q$ and $U$ images is  7.2\,$\mu$Jy, about 30\% lower than Stokes $I$. 
Figure~\ref{fig:gmrt_pol} shows the polarized intensity mosaic of the same region shown in Figure~\ref{fig:gmrt_total}.
Numerous polarized sources are seen.  Virtually all of the stronger classical double radio galaxies are visible. 
To explore statistically the potential presence of polarized signals from the fainter population we undertook a 
stacking analysis.  
For all sources in six bins of total flux density ranging from 60\,$\mu$Jy to 10 mJy, 
median stacked images in Stokes $I$ and polarized intensity were created.
Sample results are shown in Figure~\ref{fig:polstack}.
There is evidence of polarized signals down to Stokes I flux densities of 200\,$\mu$Jy -- well into the
flux density regime of star forming galaxies. 


\begin{figure}
\begin{center}
\begin{tabular}{p{7.4cm}cp{3.6cm}}
\raisebox{-\height}{\includegraphics[width=8.2cm]{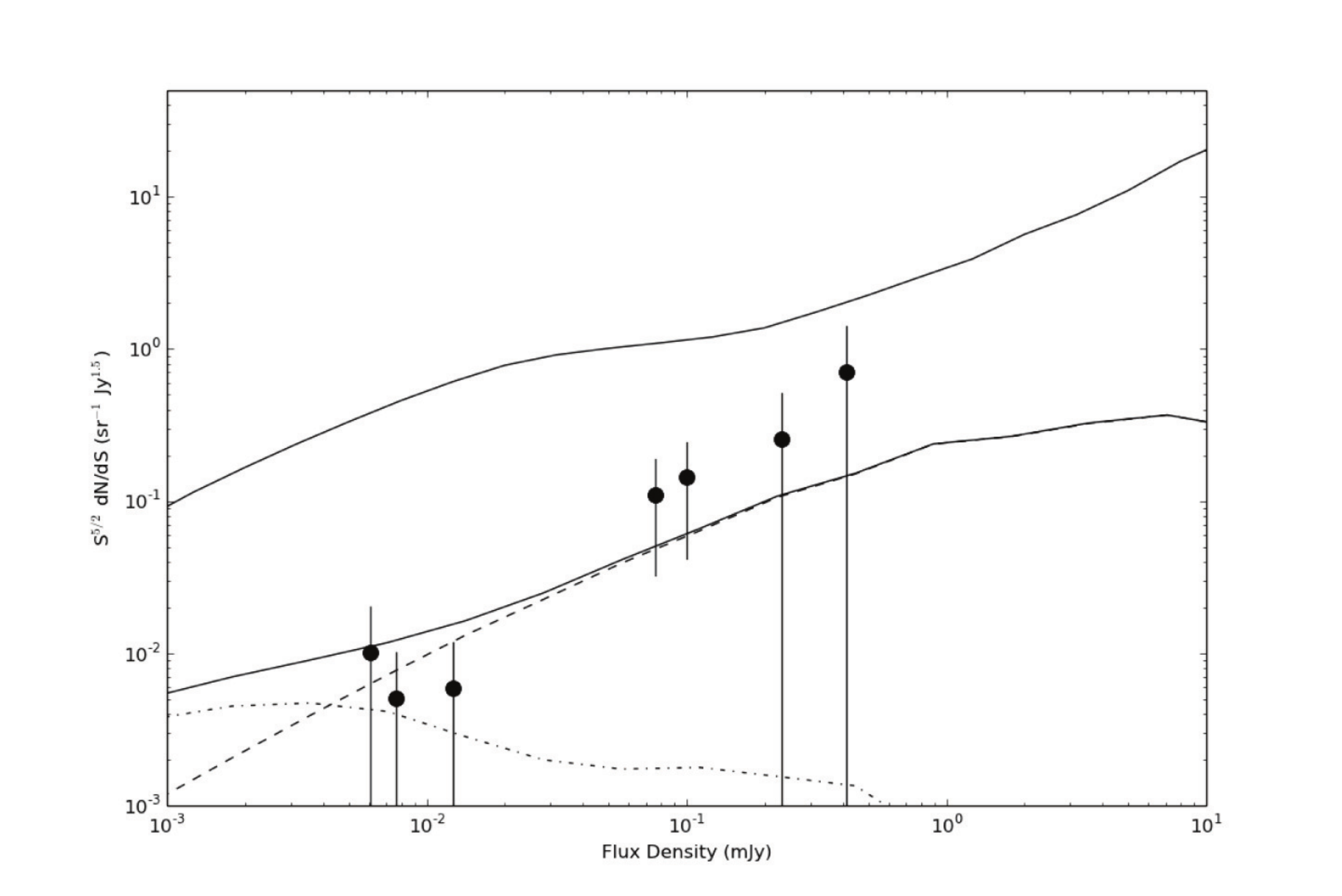}} & \quad &
\caption{Polarized source counts from the JVLA at 5 GHz.  The top solid black line
shows the total intensity counts from the SKADS simulations \citep{Wilman}.
The dashed line is the predicted 5 GHz polarized intensity counts for AGN.  The dot-dashed line
indicates the predicted polarized counts for disk galaxies.  
\label{fig:polcounts}}
\end{tabular}
\end{center}
\end{figure}

Derivation of actual median percent polarizations from the stacking analysis will require Monte Carlo simulations, since there is a 
non-linear relationship between intrinsic and observed polarized intensity in stacking  (Stil et al. 2014).
However the results suggest a decrease in median fractional polarization below Stokes I flux densities of about a mJy. 
This decrease may reflect a smaller overall fraction of polarized sources perhaps arising from a weak AGN component within star forming
galaxies or a small fractional polarization arising from galaxy disks.  
The latter is consistent with the polarization properties of galaxies predicted of \cite{Stil09} in which internal depolarization
effects from thermal plasma decreases the fractional polarization of disk emission at frequencies below a few GHz.  
In this case higher fractional polarization would occur at higher frequencies, where internal depolarization is reduced.  
Figure~\ref{fig:polcounts} show the polarized source counts from the JVLA 5 GHz images.  
The dot-dashed in the figure shows the predicted polarized counts for disk galaxies based on 
the Stil et al. polarization properties of nearby galaxies.

These preliminary results suggest that we are beginning to probe polarized emission from distant galaxies at 
$\mu$Jy  flux densities. 
Detection to deeper fractional polarization levels will require off-axis corrections 
using the wide-band AW projection \citep{Bhatnagar13} with
GMRT beam measurements which are underway \citep{Jagannathan14}.   
Beyond this, the new broad-band upgrade to the GMRT will improve sensitivity by an order of 
magnitude, allowing polarization imaging at the sub-$\mu$Jy regime, and opening up the systematic study of the magnetic 
properties of galaxies in the distant universe as a major step toward the SKA.


\appendix

\label{lastpage}
\end{document}